# KU-ISPL Language Recognition System for NIST 2015 i-Vector Machine Learning Challenge

*Suwon Shon*[1], *Seongkyu Mun*[1], *John H. L. Hansen*[2], *Hanseok Ko*[1]

[1] School of Electrical Engineering, Korea University, Seoul, South Korea
[2] Center for Robust Speech Systems (CRSS), University of Texas at Dallas, Texas, U.S.A

{swshon, skmoon}@ispl.korea.ac.kr, john.hansen@utdallas.edu, hsko@korea.ac.kr

## Abstract

In language recognition, the task of rejecting/differentiating closely spaced versus acoustically far spaced languages remains a major challenge. For confusable closely spaced languages, the system needs longer input test duration material to obtain sufficient information to distinguish between languages. Alternatively, if languages are distinct and not acoustically/linguistically similar to others, duration is not a sufficient remedy. The solution proposed here is to explore duration distribution analysis for near/far languages based on the Language Recognition i-Vector Machine Learning Challenge 2015 (LRiMLC15) database. Using this knowledge, we propose a likelihood ratio based fusion approach that leveraged both score and duration information. The experimental results show that the use of duration and score fusion improves language recognition performance by 5% relative in LRiMLC15 cost.

**Index Terms**: i-vector, language recognition, duration, GMM, DNN, fusion

## 1. Introduction

In recent years, the i-Vector based representation of speech has been used for state-of-the-art modeling with significant progress in the area of speaker and language recognition[1]–[6]. Due to this impact, National Institute of Standards and Technology (NIST) coordinated the Language Recognition i-Vector Machine Learning Challenge 2015 (LRiMLC15) which was based on the i-Vector paradigm, which differed from previous regular NIST Language Recognition Evaluations that focused on the original audio files for train and test. By providing only the i-Vector data directly, it encouraged participations from researchers not only in the audio processing field but also from the machine learning field. Participants in this evaluation uploaded their solution results to an online leader board in real-time for scoring.

The development data provided by NIST for LRiMLC15 was unlabeled. Thus, it opened up a range of opportunities for exploring new ideas with unlabeled data. Another feature of the data provided by NIST is the segment duration. Every data stream has duration information from the original input segment. Therefore, the use of duration information could potentially improve performance for evaluation. Several studies have verified that performance improvement of i-Vector system using duration by calibrating the score when there are duration mismatched conditions [7]–[11]. The statistical characteristic of i-Vector is changed by duration of the original input segment. Thus, for text-independent speaker/language recognition tasks, duration is important information for calibrating the scores. To calibrate a score by compensating for the duration mismatched condition, a Quality Measure Function (QMF) based score calibration was considered as an important process [7]–[9].

In this study, we propose an alternative approach that utilizes duration as one of the metrics for fusion to reflect the phonetic characteristics of each language. Typically, an i-Vector paradigm is used in both speaker recognition and language recognition. However, the information that the recognition system needs is different between that of speaker versus language. The main purpose of speaker recognition is to obtain speaker's phonetic characteristics from input speaker's segment regardless of their language. However, the main purpose of language recognition is to obtain the phonetic characteristics of each language while suppressing the individual speaker's phonetic characteristics. Therefore, for robust language recognition, the system should model the phonetic characteristics of each language and emphasize the differences among them for effective recognition. So, training i-Vectors for each language is usually modeled by Gaussian Mixture Model (GMM) or Deep Neural Network (DNN) [12]. Finally, the score is evaluated between the language model and input segment. We address this point for performance improvement by using duration information of input segments for robust scoring.

In language identification, a cluster based hierarchical approach has been studied previously [13], [14]. In these studies, similarities between each language were measured by a distance assessment and clustered by its similarity. Previous work on perceptual as well as acoustic dialect and language distance assessment has resulted in several effective solutions [15]. Though each language has its own phonetic rules, some languages (i.e., dialects) are very similar to other languages from the perspective of phonetic characteristics. Thus, any similar language models will lead to confusion in language recognition, and therefore a longer duration test segment is potentially needed to effectively distinguish the unique language pair. In the score domain, we validate this difference of similarity between each language by associating the duration using the NIST LRiMLC15 database. To reflect the difference for each language, we propose a likelihood ratio based fusion approach using both score and duration information. We validate the performance improvement of the proposed approach using i-Vectors from NIST LRiMLC15 and the online scoreboard of NIST LRiMLC15.

## 2. Baseline

All i-Vectors and a baseline system for LRiMLC15 are provided by NIST. From a total variability perspective, a speaker utterance can be defined by both speaker and channel variability in a single space, (i.e., total variability space), as shown in the equation below,

$$\mathbf{M} = \mathbf{m} + \mathbf{T}\omega \qquad (1)$$

where **M** represents the speaker utterance in the GMM supervector space, **m** denotes the speaker and channel independent GMM supervector from GMM-Universal Background Model (UBM). **T** is a total variability matrix which is a rectangular matrix of low rank and ω is an i-vector. NIST provided three kinds of i-vector databases as summarized below.

- Development i-Vectors: a total of 6351 i-Vectors are used for estimating the UBM and i-Vector extractor. The database have no language label.
- Training i-Vectors: a total of 15,000 i-Vectors are used for enrollment of the 50 languages. The database has labels and each language has 300 i-Vectors.
- Testing i-Vectors: a total of 6500 i-Vectors are used for test with no language label.

The database has duration information for each i-Vector as well. For the baseline system, NIST provided a simple recipe using cosine distance scoring. All i-Vectors are centered and whitened by a global mean and covariance of the development i-Vectors. Next, all i-Vectors are projected on to a unit sphere. Training i-Vectors for each language are averaged and then projected onto the average-language i-Vector in the unit sphere again. Finally, the cosine score is obtained by the inner product between all the average-language i-Vectors and testing i-Vectors.

## 3. Subsystems

Fusion of subsystems in the past have shown good performance in speaker recognition challenges. Here, we also fuse different systems to improve results. Two kinds of state-of-art posteriors [12] are used for language recognition, one employing a GMM and the other using a DNN.

### 3.1. GMM posterior based LR system

For the GMM posterior based LR system, we use a general GMM-UBM framework. Using the development i-Vector data, a universal background i-vector model $\lambda_{bkg}$ is trained with a GMM algorithm and each i-vector model $\lambda_{lang}$ for each language is created from $\lambda_{bkg}$ using maximum a posteriori (MAP) estimation. For each input i-Vector, the score can be calculated by the likelihood ratio between the universal background language model $\lambda_{bkg}$ and each language model $\lambda_{lang}$.

### 3.2. DNN posterior based LR system

For the DNN posterior based LR system, we follow a successive method using DNN posterior for SR based on Deep Belief Networks (DBN) [16]. At first, i-Vectors for DNN input are prepared from the train database. Next, a Universal DBN (UDBN) is trained using unlabeled speaker i-Vectors. After the UDBN is trained in an unsupervised manner, a label layer is added on top of the network and a stochastic gradient descent backpropagation is carried out for overall fine tuning of the DNN.

## 4. Proposed duration and score fusion

In this section, we introduce duration and score fusion based on likelihood ratio for reflecting phonetic characteristics of the languages. We assume that every language has its own phonetic characteristics, however some languages are similar to each other. In such cases, the duration would be much more important to distinguish between similar languages.

In the case of speaker recognition, duration is important for obtaining sufficient information of unique phonemes. The number of unique phonemes increases by duration exponentially, so that the statistical characteristics of the i-Vectors is altered by duration [9]. Thus, duration is a significant value in determining a score for the input test utterance.

However, in the case of language recognition, duration is significant from a different perspective compared with speaker recognition. Suppose there is a limited number of languages for a system to recognize. If one language is very unique and not similar (i.e., far acoustic distance) to other languages, duration is not important because the phonetic characteristics of the smaller test set are sufficiently different. However, if the phonetic characteristics of a language are very similar to other languages in the system, duration is much more important to leverage information of the input segment in distinguishing the differences between languages. It would be possible to explore this through a score by duration distribution analysis.

The score can be calculated by a GMM posterior as noted in Sec. 3.1 and specific parameters for obtaining GMM score is described in Sec. 5. Leave-one-out Cross-validation method is used for obtaining scores among the train i-Vectors, because the training i-Vectors only have language labels. For one i-Vector, we can obtain the 1 target score and the 49 non-target language scores. In total, we obtain 15,000 target scores and 735,000 non-target scores and all duration is converted to a logarithmic scale. Fig. 1 shows a scatter plot of the score duration distribution of target scores and non-target scores of all language train i-Vectors. In general, the shorter duration i-Vector will be difficult to classify as target or non-target. For the longer duration, the scores are easier to classify as target or not. At this step, by using a calibration approach based on using duration, the target and non-target score can be calibrated for an overall optimal decision threshold as previous studies [7]–[9]

However, if we look at the specific languages, for example, Arabic and Romanian as below Fig. 2 and 3 respectively, it is clear that the score distribution is totally different for these languages. For the Romanian language, the target scores show constant high value regardless of duration. For the Arabic language, a long test duration is necessary to achieve the same consistent high score.

From these analyses, we can verify that some languages which are more confusable require longer test duration than languages which are more acoustically separable. By utilizing this difference of score distribution, it could achieve a robust score through fusion of duration and score values for each language. One possible reason for success of this approach is that in language recognition, we typically have a sufficient size enrollment (i.e., training i-Vectors) database for each language to model the score densities, while, for speaker recognition, the number of enrollment entries in the database is small for each speaker. The other reason is that there is a limited number of classes (language) to recognize, so it is possible to establish the target class score density models in advance.

Suppose $x_l$ is the score of $l$ language class using an input speech utterance $x$ of $d$ duration. The joint densities of score and duration given the target and non-target classes are $f(x_l, d|\lambda_{tar})$ and $f(x_l, d|\lambda_{non})$ respectively. Here, a GMM is successfully used to estimate the arbitrary densities in [17], [18]. Thus, the estimates of the joint densities of score and duration are:

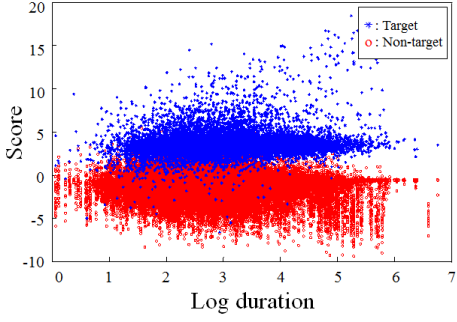

Figure 1: *Score by Duration distribution on all-languages using GMM posterior score*

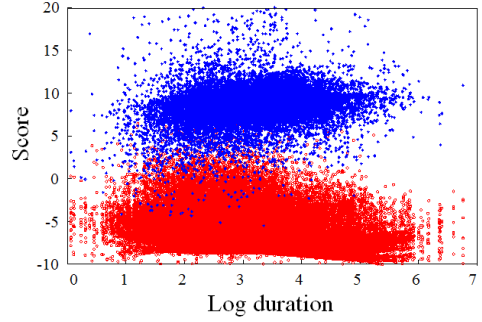

Figure 4: *Proposed fusion score by duration distribution on all-languages using GMM posterior*

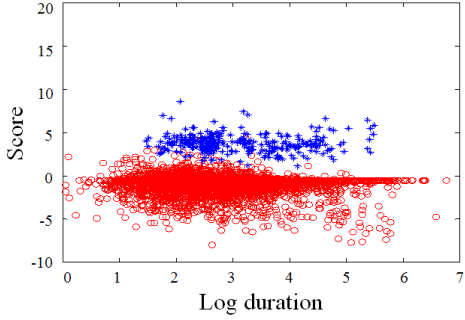

Figure 2: *Score by Duration distribution on Romanian language using GMM posterior*

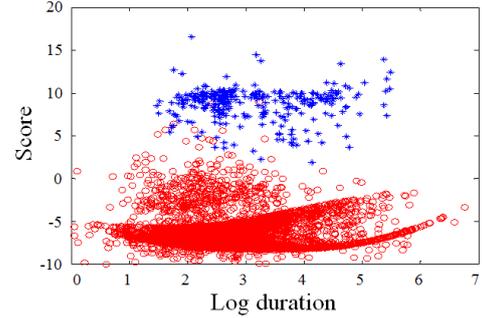

Figure 5: *Proposed fusion score by duration distribution on Romanian language using GMM posterior*

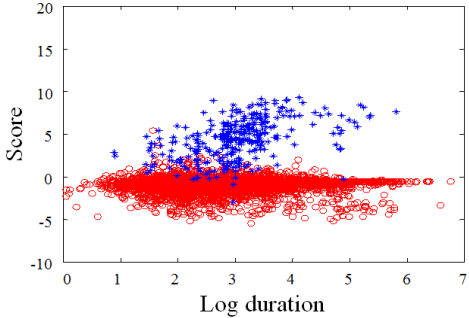

Figure 3: *Score by Duration distribution on Arabic language using GMM posterior*

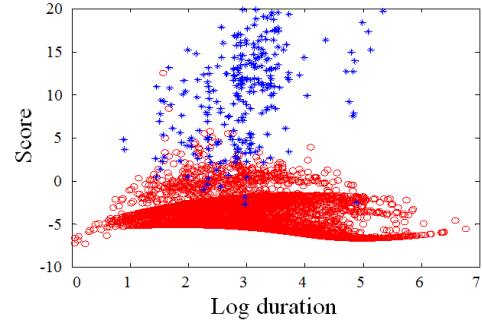

Figure 6: *Proposed fusion score by duration distribution on Arabic language using GMM posterior*

$$f(x_l, d \mid \lambda_{tar}) = \sum_{i=1}^{C} w_{tar,i} g(x_l, d \mid \mu_{tar,i}, \Sigma_{tar,i}) \quad (2)$$

$$f(x_l, d \mid \lambda_{non}) = \sum_{i=1}^{C} w_{non,i} g(x_l, d \mid \mu_{non,i}, \Sigma_{non,i}) \quad (3)$$

where $\lambda_{tar}$ and $\lambda_{non}$ is target and non-target score GMM model, $\mu$ and $\Sigma$ and $w$ are the mean, covariance and weight of the GMM respectively. $C$ is total number of components. The likelihood ratio is then,

$$LR(x_l, d) = \frac{f(x_l, d \mid \lambda_{tar})}{f(x_l, d \mid \lambda_{non})} \quad (4)$$

and the system output is likelihood ratio vector as below,

$$\text{system output} = [LR(x_{l=1}, d), LR(x_{l=2}, d), ..., LR(x_{l=L}, d)] \quad (5)$$

where total number of language $L$=50. We assume input $x$ to $l$ language class if $LR(x_l, d) > \eta$, and out of set if all $LR(\cdot) < \eta$ where $\eta$ is threshold. For reflecting each of the language characteristics, the joint density can be revised using $\lambda^l_{tar}$ and $\lambda^l_{non}$ which are score models of $l$ language class as follows,

$$f(x_l, d \mid \lambda^l_{tar}) = \sum_{i=1}^{C} w^l_{tar,i} g(x_l, d \mid \mu^l_{tar,i}, \Sigma^l_{tar,i}) \quad (6)$$

$$f(x_l, d \mid \lambda^l_{non}) = \sum_{i=1}^{C} w^l_{non,i} g(x_l, d \mid \mu^l_{non,i}, \Sigma^l_{non,i}) \quad (7)$$

$$LR_l(x_l, d) = \frac{f(x_l, d \mid \lambda^l_{tar})}{f(x_l, d \mid \lambda^l_{non})} \quad (8)$$

$$\text{system output} = [LR_{l=1}(x_{l=1}, d), LR_{l=2}(x_{l=2}, d)..., LR_{l=50}(x_{l=L}, d)] \quad (9)$$

For estimating the GMM parameters of $\lambda^l_{tar}$ and $\lambda^l_{non}$ for each language, we employ a GMM-UBM paradigm [19]. Universal target score model can be estimated by target score of all languages. Next, the each language target score model can be adapted by MAP as shown in Fig. 7. This process is also done to obtain the non-target score model. The number of optimal GMM component, can be estimated by a minimum message length criterion which is proposed in [20] for GMM fitting algorithm.

We can assess the effect of the proposed approach as Fig. 4 to 6. Figure 4 shows the target and non-target scores distribution of all languages. We can check the distance between target and non-target scores, which are clearly farther than before. For the each language's score distribution as Fig. 5 and 6, the non-target scores decrease while the target scores increase compare to the original score distribution seen in Fig. 3 and 4. This changes will affect the false positive rate of the system significantly. We can verify these improvements with a Detection Error Trade-off (DET) curve in Fig. 8. From these analyses, using duration for fusion to reflect the phonetic characteristics of each language shows promising performance improvement in language recognition.

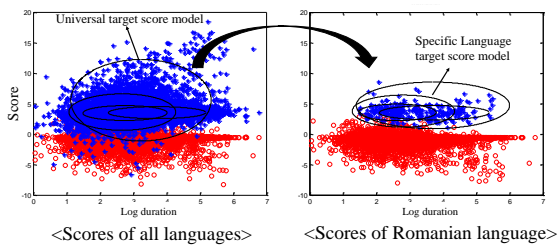

Figure 7: *Estimation of target score density model*

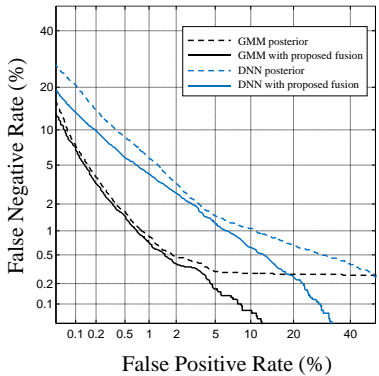

Figure 8: *Detection Error Trade-off curve of GMM and DNN posterior systems with proposed fusion*

## 5. KU-ISPL system

In this section, we describe the specification of Korea University – Intelligent Signal Processing Laboratory (KU-ISPL) system used in NIST LRiMLC15.

At first, the dimension of the i-Vectors was reduced from 400 to 49 using Linear Discriminant Analysis (LDA). The LDA transformation matrix is calculated by training i-Vectors which have language labels. The scores were calculated by both GMM and DNN posteriors as discussed in Sec. 3. For the GMM, 49 dimension i-Vector is trained with a 64 component GMM. The DNN used here consists of an input layer with 49 nodes, 2 hidden layers with 600 nodes and an output layer with 50 nodes. The sigmoid and softmax are the activation function of all hidden layers and top label layer, respectively.

The proposed score and duration fusion was done using both GMM and DNN posterior scores. The optimal number of GMM components was 4 for both universal target and non-target score model by minimum message length criterion [20]. The two subsystem was fused based on the linear model [21]. Next, quality calibration was done based on QMF [11] and suitable threshold $\eta$ is determined based on training i-Vectors database

## 6. Performance assessment

The performance was assessed using the progress set of the test i-Vectors. A progress set comprised of randomly selected 30% among 6500 testing i-Vectors for preventing empirical optimization of recognition system. The score metric is defined in LRiMLC15 as follows [22],

$$Cost = \frac{(1-P_{oos})}{n} * \sum_{k}^{n} P_{error}(k) + P_{oos} * P_{error}(oos) \quad (10)$$

where $P_{error}$=(#errors_class_k / #trials_class_k), $n$=50, $P_{oos}$=0.23 and *oos* for "out of set". All cost values were calculated in the NIST server by uploading the estimated language labels of test i-Vectors. For validating the proposed approach, costs were compared between the systems that applied the proposed approach and not. Table 1 shows the costs of several systems in our study. The baseline represents the system described in Sec. 2. The GMM and DNN reflect subsystems described in Sec. 3. Fusion stands for the system described in Sec 5.

For a single subsystem, the GMM subsystem shows better performance than other systems such as baseline and DNN subsystem. The systems with the proposed duration and score fusion approach shows overall better performance than systems without the proposed approach in both subsystems and fusion systems.

Table 1. *Performance measurements by LRiMLC15 cost of language recognition systems.*

| Language Recognition Systems | LRiMLC Cost (smaller=better) |
|---|---|
| Baseline | 39.590 |
| GMM subsystem (*A*) | 28.692 |
| DNN subsystem (*B*) | 32.538 |
| Fusion of *A* and *B* | 25.744 |
| GMM subsystem with proposed (*C*) | 26.974 |
| DNN subsystem with proposed (*D*) | 30.103 |
| Fusion of *C* and *D* | **24.256** |

## 7. Conclusion

This study has explored score and duration fusion for language recognition. Though score calibration using duration has been proposed in recent studies, the analysis of similarity between language classes based on duration has not been explored. The purpose of this study has been to show that score distribution by duration is discriminative between each language. Based on original phonetic characteristics, each language might be similar to other languages or not. Using these differences, we proposed a likelihood ratio based fusion approach which leverages score and duration. From analysis and online score results from the NIST LRiMLC15 leader board, we have validated the performance improvement of the fused GMM and DNN language recognition system. Therefore, it is clear that using duration for fusion with the score on each language is effective for improving overall language recognition performance.